\documentclass[letterpaper,twocolumn,10pt]{article}
\usepackage{usenix,epsfig,endnotes}

\newcommand{\MOBSEC}{\textit{Mobsec Analytika}}

\def\experimentDurationDays{150} %

\def\androZooMatches{12,033,563} %

\def\collectedApps{973,014}

\def\analyzedApps{170,550}
\def\analyzedSuccessApps{170,516}
\def\analyzedFailedApps{34}

\def\pkgNotMatching{103}
\def\pkgNotMatchingStartswith{41}

\def\matchAllTee{4}
\def\matchAllButConfPrompt{3,268}
\def\matchAllButConfPromptPercent{1.9\%}
\def\matchAtLeastOneTee{74,829}
\def\matchAtLeastOneTeePercent{43.9\%}
\def\matchNoTee{95,687}
\def\matchNoTeePercent{56.1\%}

\def\matchCryptoImports{167,420}
\def\matchCryptoImportsPercent{98.2\%}

\def\matchKeystoreTotal{54,540}
\def\matchKeystoreTotalPercent{32.0\%}

\def\matchBiometricsTotal{11,674}
\def\matchBiometricsTotalPercent{6.8\%}

\def\matchDrmTotal{38,743}
\def\matchDrmTotalPercent{22.7\%}

\def\matchConfPromptTotal{4}
\def\matchConfPromptTotalPercent{0.0\%}

\def\matchCryptoImports{167,600}
\def\matchCryptoImportsPercent{98.6\%}
\def\matchCryptoLibJSE{167,599}
\def\matchCryptoLibAndX{3,706}
\def\matchCryptoLibBoCa{8,306}
\def\matchCryptoLibNimDS{4,854}
\def\matchCryptoLibSprSec{13}
\def\matchCryptoLibJasypt{7}
\def\matchCryptoLibJNCryptor{102}
\def\matchCryptoLibAWSKMS{1,760}
\def\matchCryptoLibGoTink{5,634}
\def\matchCryptoLibApaComm{1}
\def\matchCryptoLibApaTuw{0}

\def\matchAtLeastOneNativeLib{410}
\def\matchCryptoCryptoPP{32}
\def\matchCryptoBotan{0}
\def\matchCryptoLibG{0}
\def\matchCryptoOpenSSL{401}
\def\matchCryptoGnuTLS{0}
\def\matchCryptoSodium{30}
\def\matchCryptoNettle{0}
\def\matchCryptoWolfSSL{4}

\def\matchObfuscatedAppsWithMatchPercent{23.0\%}

\def\elapsedTimeMedian{23.23}
\def\elapsedTimeMin{0.21}
\def\elapsedTimeMax{814.76}

\def\matchInlibApps{70,611}
\def\matchInlibAppsPercent{41.5\%}
\def\matchInlibAppsWithMatchPercent{94.4\%}

\def\inlibMatchesPercent{91.6\%}

\def\inlibMeanLibraryCount{1.94}
\def\inlibMedianLibraryCount{1.0}

\def\inlibKeystorePackages{5,661}
\def\inlibDrmPackages{2,583}
\def\inlibBiometricsPackages{638}
\def\inlibProtconfPackages{2}

\def\inlibMatchKeystorePercent{68.0\%}

\def\inlibMatchBiometricsPercent{15.6\%}

\def\inlibMatchDrmPercent{53.9\%}
\def\inlibMatchProtconf{4}
\def\inlibMatchProtconfPercent{0.0\%}

\def\matchInmainApps{11,364}
\def\matchInmainAppsPercent{6.7\%}
\def\matchInmainAppsWithMatchPercent{15.2\%}

\def\matchInmainOnlyAppsWithMatchPercent{5.0\%}

\def\inmainMatchKeystorePercent{87.5\%}

\def\inmainMatchBiometricsPercent{8.5\%}

\def\inmainMatchDrmPercent{14.0\%}

\newcommand{\tblsizeappendix}{\tiny}

\usepackage[utf8]{inputenc}
\usepackage{csquotes}
\usepackage{chngcntr}  %
\usepackage{float}  %
\usepackage{enumitem}
\setlist{nosep,before=\vspace{.5\baselineskip},after=\vspace{.5\baselineskip}}
\usepackage{listings}  %

\usepackage{xurl}

\usepackage{arydshln}

\usepackage[prependcaption]{todonotes}

\usepackage{amsmath}

\usepackage{csvsimple}

\usepackage{subcaption}

\usepackage[upgrade=true]{acro}[=v3]
\acsetup{
  use-id-as-short,
  single = true,
  patch/longtable=false  %
}

\DeclareAcronym{SoK} {
    long    =   Systematization of Knowledge,
}

\DeclareAcronym{UI} {
    long    =   User Interface,
}

\DeclareAcronym{TUI} {
    long    =   Trusted User Interface,
}

\DeclareAcronym{a11y} {
    long    =   Accessibility
}

\DeclareAcronym{PoC} {
    long    =   Proof of Concept
}

\DeclareAcronym{AOSP} {
    long    =   Android Open-Source Project
}

\DeclareAcronym{DSL} {
    long    =   Domain Specific language
}

\DeclareAcronym{TCB} {
    long    =   Trusted Computing Base
}

\DeclareAcronym{RoT} {
    long    =   Root of Trust,
    long-plural-form  = Roots of Trust
}

\DeclareAcronym{TPM} {
    long    =   Trusted Platform Module
}

\DeclareAcronym{IPC} {
    long    =   inter-process communication
}

\DeclareAcronym{TEE} {
    long    =   Trusted Execution Environment
}

\DeclareAcronym{SGX} {
    short   =   SGX,
    alt     =   Intel SGX,
    long    =   Intel Software Guard Extensions,
}

\DeclareAcronym{CSR} {
    long    =   Control and Status Register,
}

\DeclareAcronym{ROP} {
    long    =   Return-Oriented Programming
}

\DeclareAcronym{RTOS} {
    long    =   Real-Time Operating System,
}

\DeclareAcronym{ISA} {
    long    =   Instruction Set Architecture
}

\DeclareAcronym{SM} {
    long    =   Secure Monitor
}

\DeclareAcronym{PMP} {
    long    =   Physical Memory Protection
}

\DeclareAcronym{IoT} {
    long    =   Internet-of-Things
}

\DeclareAcronym{CFI} {
    long    =   Control Flow Integrity
}

\DeclareAcronym{ASLR} {
    long    =   Address Space Layout Randomization
}

\DeclareAcronym{MMU} {
    long    =   Memory Management Unit
}

\DeclareAcronym{DoS} {
    long    =   Denial-of-Service
}

\DeclareAcronym{SSH} {
    long    =   Secure Shell
}

\DeclareAcronym{JSON} {
    long    =   JavaScript Object Notation
}

\DeclareAcronym{UART} {
    long    =   Universal Asynchronous Receiver-Transmitter
}

\DeclareAcronym{TZPC} {
    long    =   ARM TrustZone Protection Controller
}

\DeclareAcronym{SMC} {
    long    =   Secure Monitor Call
}

\DeclareAcronym{HPC} {
    long    =   Hardware Performance Counter
}

\DeclareAcronym{RT} {
    long    =   Runtime
}

\DeclareAcronym{OS} {
    long    = Operating System
}

\DeclareAcronym{UDS} {
    long    =   Unique Device Secret
}

\DeclareAcronym{KDF} {
    long    =   Key Derivation Function
}

\DeclareAcronym{MQTT} {
    long    =   Message Queue Telemetry Transport
}

\DeclareAcronym{ECDSA} {
    long    =   Elliptic Curve Digital Signature Algorithm
}

\DeclareAcronym{LoC} {
    long    =   Lines of Code
}

\DeclareAcronym{DICE} {
    long    =   Device Identifier Composition Engine
}

\DeclareAcronym{ecall} {
    long    =   environment call
}

\DeclareAcronym{ePMP} {
    long    =   Enhanced Physical Memory Protection
}

\DeclareAcronym{API} {
    long    =   Application Programming Interface,
}

\DeclareAcronym{HAL} {
    long    =   Hardware Abstraction Layer
}

\DeclareAcronym{M-Mode} {
    long    =   Machine Mode
}

\DeclareAcronym{S-Mode} {
    long    =   Supervisor Mode
}

\DeclareAcronym{U-Mode} {
    long    =   User Mode
}

\DeclareAcronym{H-Mode} {
    long    =   Hypervisor Mode
}

\DeclareAcronym{SDK} {
    long    =   Software Development Kit
}

\DeclareAcronym{RISCoT} {
    long    =   RISC-V of Trust
}

\DeclareAcronym{LiME} {
    long    =   Linux Memory Extractor
}

\DeclareAcronym{VM} {
    long    =   Virtual Machine,
}

\DeclareAcronym{eapp} {
    long    =   enclave application,
}

\DeclareAcronym{DRM} {
    long    =   Digital Rights Management
}

\DeclareAcronym{APK} {
    long    =   Android Package
}

\DeclareAcronym{DEX} {
    long    =   Dalvik Executable
}

\begin{document}

\date{}

\title{\Large \bf A Large-Scale Study on the Prevalence and Usage of TEE-based Features on Android}

\author{
{\rm Davide Bove}\\
FAU Erlangen-Nürnberg, Germany\\
{\rm\href{mailto:davide.bove+arxiv@fau.de}{davide.bove@fau.de}}
}

\maketitle

\begin{abstract}
In the realm of mobile security, where OS-based protections have proven insufficient against robust attackers, Trusted Execution Environments (TEEs) have emerged as a hardware-based security technology.
Despite the industry's persistence in advancing TEE technology, the impact on end users and developers remains largely unexplored. 
This study addresses this gap by conducting a large-scale analysis of TEE utilization in Android applications, focusing on the key areas of cryptography, digital rights management, biometric authentication, and secure dialogs.

To facilitate our extensive analysis, we introduce Mobsec Analytika, a framework tailored for large-scale app examinations, which we make available to the research community.
Through \analyzedApps{} popular Android apps, our analysis illuminates the implementation of TEE-related features and their contextual usage.

Our findings reveal that TEE features are predominantly utilized indirectly through third-party libraries, with only \matchInmainAppsPercent{} of apps directly invoking the APIs.
Moreover, the study reveals the underutilization of the recent TEE-based UI feature Protected Confirmation.
\end{abstract}

\section{Introduction}

In Mobile Security, recent work has shown that \ac{OS}-based protections are inadequate to protect devices against strong attackers~\cite{DBLP:journals/ese/Mazuera-RozoBLR19,DBLP:conf/asiaccs/Bove22}.
Therefore, vendors implement features based on \acfp{TEE}, a hardware-based architectural protection that promises to protect sensitive data in the presence of a compromised \ac{OS} and against strong attackers.
Even though research has also shown several problems with \ac{TEE} solutions in the past, both research and industry push on with this technology, adding functionality based on it to recent devices.

While the goal to improve the security of devices is surely a noble cause, the security community has failed to check the impact of their new developments on the actual users of these protections: the consumers and developers of mobile apps and devices.
To our knowledge, there are only a handful of evaluations on how mobile security research and developments actually affect the user experience (e.g., \cite{DBLP:conf/uss/ImranFICB22, DBLP:phd/dnb/Kraus17}).
This work raises the question if and to what extent users get in touch with such technologies, and provides novel insights in how to improve the situation.

The impact of security measures is difficult to measure, and there is barely publicly available data of mobile app security.
It is increasingly more difficult to collect large numbers of Android apps, as Google has limited \ac{API} access to app metadata and content, and web crawlers are regularly made ineffective by changes to the Play Store website.
Moreover, obfuscation of apps further complicates static analysis, and the available tools, especially from academia, are quickly deprecated or not applicable.
Therefore, we enter this research area with really limited data and related work for comparison.

In this work, we want to quantify how \acp{TEE} are used on mobile devices.
For this, we focus on the following research questions:

\begin{itemize}
  \item \textbf{How many apps use \acs{TEE}-based security features?}
  
    We look at four specific \acp{API} backed by \ac{TEE} that developers may use:
    \begin{itemize}
      \item Cryptography
      \item \acl{DRM}
      \item Biometric Authentication
      \item Protected Confirmation
    \end{itemize}

    \item \textbf{Where and in which context are these features used?} 
    
      We want to know how the corresponding \ac{API} is used.
      Do developers actively implement it, or are the calls part of an included software library?
\end{itemize}

For this work, we collected \collectedApps{} apps over a study period of \experimentDurationDays{} days (April 2023 to September 2023).
Using static analysis tools, we extract function calls and class invocations to establish if an \ac{API} is being used inside an app.
Apart from their general \ac{TEE} usage, we also collect general information about the apps, such as their popularity derived from the download count and how they are categorized inside the Google Play Store.
In a last step, we aggregate the results using the \textit{pandas} data analysis framework, which produces several artifacts that we present here.

To summarize, \textbf{our contributions} are:
\begin{enumerate}
  \item We conduct the first large-scale analysis of \ac{TEE} utilization in Android applications.
    We delve into four distinct \acp{API} and examine their usage within \analyzedApps{} Android apps.
    Provided with all details of our methodology, readers may reproduce our analysis or extend it for future research.
    
  \item We present and publish our analysis framework \MOBSEC{}, which combines several static analysis techniques, and is tailored for large-scale app experiments by security researchers and professionals.
\end{enumerate}

The remaining work is structured as follows:
We give a brief introduction to the basics of app analysis and \ac{TEE} in \autoref{forgotee:background}.
We describe the methodology of our large-scale analysis in \autoref{forgotee:methodology}.
Then we present insights into our analysis in \autoref{forgotee:analyzing} and present our results in \autoref{forgotee:results}.
We discuss related work in \autoref{forgotee:related-work} and present potential further refinements in \autoref{forgotee:conclusion}.

\section{Background} \label{forgotee:background}
This section provides the background to Android security features, the structure of Android apps and general information about the \acp{API} that we target in our experiments.
We also give an overview of \aclp{TEE} in the context of Android.

\subsection{Android security}
Android is an \ac{OS} based on a modified Linux kernel that enforces strong isolation between apps.
This isolation is achieved with the Android Sandbox, which prevents an application from accessing the code and data of other apps.
With a permission-based model, apps can request access to parts of the system or device resources.
Developers use the Android \ac{SDK} to access various \acp{API} that grant access to resources or allow interacting with other apps.
Android uses multiple abstraction layers between apps and hardware, such that the code to access resources is decoupled from the actual driver that performs system operations.
Therefore, apps written once can run on almost all Android devices on the market.

Apps are packaged in \acf{APK} files, which contain all the code and resources.
They can be written in Java, Kotlin or using native interfaces with C or C++.
The \ac{APK} contains the compiled byte code as \ac{DEX} files and all app data, such as texts, images, and configurations.
It also includes the \texttt{AndroidManifest} file, which contains meta-information required to install the app, such as required permissions, the (unique) package name and any hardware or software requirements of the app.
Finally, the package may also include native code software libraries.

An \ac{APK} file is an archive which contains several other files.
One of these files is \texttt{classes.dex}, which contains the compiled bytecode for the app and is often the main file used for static code analysis.
There can be several of these files, which are then named \texttt{classes1.dex}, \texttt{classes2.dex} and so on.
Bytecode can usually be decompiled into (human-readable) Java code, as the file format preserves much information about the program, such as strings, constants, class method names and more.
Apps can also contain native code libraries, which are typically found in the app's resources as shared object (\texttt{.so}) files.
Another important file in Android apps is the \texttt{AndroidManifest} file, which contains meta-information required to install the app, such as required permissions, the (unique) package name and any hardware or software requirements of the app.
It also contains all references to the app's services, screens and more.
As apps need to declare all permissions in the manifest, it is often used as a starting point to estimate the expected app behavior and functionality.
Since \ac{DEX} files contain very detailed internal information about the code, such as function names and identifiers, they can be easily reversed to get the original source code.
This process of decompilation is the basis for most static analysis tools for Android apps.
Popular app analysis tools for \acp{APK} are \texttt{jadx}, a Java bytecode decompiler, and \texttt{apktool}, a software that converts the byte code into the intermediate \texttt{smali} format which can more easily be analyzed and modified.

\subsection{Trusted Execution on Android}
\begin{figure}[htbp]
    \centering
    \includegraphics[width=\linewidth]{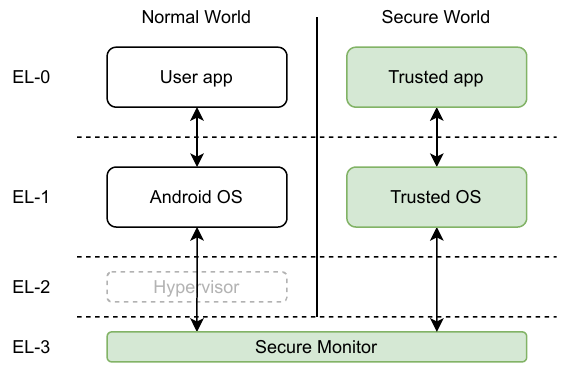}
    \caption{The TrustZone architecture on Android}
    \label{fig:forgotee-trustzone}
\end{figure}

Almost all modern mobile devices recently have adopted \acp{TEE} into their system design.
\Ac{TEE} is a security architecture that aims to maintain an isolated environment on the same device and hardware.
Since Android only supports ARM and x86 architectures (both 32-bit and 64-bit versions), and there are only very few x86-based mobile devices, ARM processors are predominantly used in the mobile market~\cite{androidAndroidABIs2023}.
Therefore, the ARM TrustZone is used to implement \ac{TEE} features.
The general architecture of devices with TrustZone contains a second \ac{OS}, the Trusted \ac{OS}.
Single security features are bundled into trusted apps, so-called trustlets.
These can provide several security-related services, such as mobile payments, user authentication, and cryptographic operations.
The general architecture of TrustZone is shown in \autoref{fig:forgotee-trustzone}.

While device vendors usually provide binary blobs of \ac{TEE} implementations and apps, Android provides a free and open-source alternative to be used, an own \ac{TEE} \ac{OS} implementation called Trusty~\cite{googleTrustyTEE2022}.
Different vendors can implement trustlets differently, so we focus our research on the interface between normal \ac{OS} and \ac{TEE}.
In the following, we present the relevant \acp{API} that are analyzed in our experiments.

\subsection{KeyStore}
The Keystore \ac{API} is responsible for cryptographic operations: it can encrypt and decrypt data, create signatures and provides a persistent storage for cryptographic keys.
It is designed such that sensitive cryptographic material (e.g., private keys and secrets) are not handled in the regular context of the \ac{OS}, but in the TrustZone, which has its own main memory and execution context.
Therefore, everything is processed in the isolated \ac{TEE} context and no sensitive data is ever passed to the \ac{OS} or the apps.
Keys are bound to an app, so no app can access or use the encryption keys of other apps.
The \ac{API} communicates with the \texttt{keymaster} trustlet, which performs all critical operations~\cite{DBLP:conf/esorics/SabtT16, mayrhofer2019}.
Apps can only access their own keys, and the export of private keys is not supported.
On devices with Android 9 and newer, the new StrongBox technology may be available on the device.
It usually consists of an additional secure coprocessor, so the actual private keys might be saved on the storage of the secure chip.
If not the case, the keys are usually encrypted and saved on the normal file system under \texttt{/data/misc/keystore}.
It takes several \ac{API} calls to safely use the Android Keystore in apps, and security professionals have demonstrated the pitfalls for developers in using it~\cite{BPF2019}.
The \ac{API} with its \ac{TEE} counterpart have existed since Android 4.3 (2013).

\subsection{\acl{DRM}}
\ac{DRM} combines techniques to protect intellectual property from being copied or manipulated.
It is used by apps that would like to display images or play audio protected by copyright, such as audio and video streaming services or games.
Behind the scenes, the Google company \textit{Widevine} provides the software to encode and decode media streams, both inside the hardware (called \enquote{Security Level 1}) and as a software-only solution (\enquote{Security Level 3})~\cite{DBLP:conf/sp/PatatSF22}.
The \ac{API} communicates with the \texttt{widevine} trustlet running inside the \ac{TEE}.
The relevant media components of the \ac{DRM} \ac{API} have been introduced with Android 4.3 in 2013. %

\subsection{Biometric Authentication}
Devices with appropriate fingerprint sensors or dedicated cameras can use these to authenticate the current user.
Especially for critical operations, such as device unlocking or confirmation dialogs, it can be a safe option to ask for the user to use their fingerprint or a scan of their face to proceed.
Android offers an \ac{API} to do this, using biometric information, without the need to pass this information to the app or give it access to the raw data.
Again, this process is mainly performed in the \ac{TEE} through the \texttt{fingerprint} trustlet, where the app directly accesses the fingerprint sensor of the device to retrieve the biometric information, saves it to be used exclusively from the trusted environment and returns the result of the authentication check to the requesting app.
For the face unlock feature of some devices, the interfacing service is called \texttt{faced}.
Android requires vendors to use secure hardware to implement this feature, such as a \ac{TEE} and \enquote{secure camera hardware}~\cite{googleFaceAuth2022}.
Still, both biometric methods are accessed by the same \ac{API} calls.
The Biometrics \ac{API} has been introduced with Android 9 in August 2018~\cite{androidbiometrics}.

\subsection{Protected Confirmation}
Since Android 9 the \texttt{ConfirmationPrompt} \ac{API} can be used to show a secure \ac{UI} dialog and ask the user for confirmation.
Protected Confirmation dialogs take up the whole screen and show a developer-defined text to confirm or dismiss.
The user can confirm the text on the screen by pressing the hardware buttons on the device or, if not present, use virtual buttons on the touch screen.
The screen is composed and displayed from the secure context of the ARM TrustZone, and overlaps any other screen contents until the dialog is taken care of~\cite{DBLP:conf/asiaccs/Bove22}.
The app receives a digitally signed confirmation that can be forwarded to third-party systems and which guarantees that the confirmation is legitimate.
Combined with biometric authentication, remote parties can request a confirmation from specific app users that is tamper-proof and transparent to the third-party entity.

While this feature has existed since Android 9, released in 2018, it is currently unknown how many devices actually support this feature.
Android defines what vendors must or may implement to ensure compatibility.
For Protected Confirmation, Android 9, 10 and 11 only \enquote{recommend} the implementation, while the specification changed to \enquote{strongly recommend} since Android 12~\cite{googleCompatibility12}.
Therefore, at the time of writing, the \ac{TEE} support of this feature is still optional.
Developers may use the \ac{API} \texttt{ConfirmationPrompt.isSupported} to check the availability for the current device.

\section{Methodology} \label{forgotee:methodology}
For our data collection and evaluation, we collected \collectedApps{} apps over a period of \experimentDurationDays{} days.
In the following section, we explain our methodology, which includes a data selection and collection phase and an analysis based on our framework \MOBSEC{}.

\subsection{Data Selection}
Based on the official Android documentation and knowledge of the developer \ac{SDK}, we identified four different \acp{API} that call functions or interfaces located in the \ac{TEE} part of the system.
While the implementation on the \ac{TEE} side might vary between different vendors, the Android framework is supposed to abstract from these implementations, therefore the code of apps using the \ac{API} stays the same for different devices.

In order to be able to reduce the number of apps to analyze while still retaining reproducibility of our experiment, we chose to exclude apps from our analysis.
The included apps need to match the following criteria:

\begin{itemize}
  \item The app needs to have at least {10,000} app downloads / installs in the Google Play Store. %
    This is mainly to select apps that have relevancy and impact.

    \item The app was updated recently. We chose January 1, 2020, as the earliest valid date for apps.
      Some selected \acp{API} have been introduced in the last few years, therefore having older apps in the dataset might significantly skew the results.

  \item The app is \textbf{not} a game. Games mostly use game engines and assets that are included into the package, and are therefore big and cumbersome to analyze.
    Early cherry-picked examples of popular games also yielded no significant or interesting results to the questions raised in our work.
    Games are filtered out of our data collection by \textit{category} as defined in the Play Store.
\end{itemize}

Apps in the Google Play Store have a category they were assigned to by their developers.
We use this category to filter out specific apps, as well as to build statistics on which types of apps use \ac{TEE} features in our results.
For example, apps of the category \enquote{Educational} refer to educational games, while the category \enquote{Education} is for non-game apps.
We identified 17 categories related to games, which we use for our pre-filtering.
While this attribute is not necessarily a reliable indicator for an app's content, it is the best we have at hand to determine the type of app.
For instance, the \enquote{Finance} category contains both apps related to banking and notes apps for personalized shopping lists.
The reliability of this information comes from the large amount of data we collect, which should keep the relative percentage of wrongly categorized apps low.
Regarding our dataset, we also did not analyze system apps, as our experiment focuses on third-party apps.
With the above requirements defined, we describe the app collection process.

\subsection{Data Collection}
In order to retrieve the \ac{APK} files for our analysis, we used the dataset of AndroZoo~\cite{DBLP:conf/msr/AllixBKT16}.
The project maintains a list of apps in their database.
We removed all entries which do not have the Play Store as source, and filtered the list by latest app versions.
With the resulting list, which contains around \androZooMatches{} different \acp{APK}, we randomly pick apps and crawl their respective Play Store website to check their existence on the store and to retrieve more information.
In cases where the Play Store entry does not exist anymore, we exclude the respective app from our list.
If it exists, we apply the filters described above and mark packages for further examination.
In this step, our Play Store crawler also collects relevant metadata, such as download count, version information, app category and the date of last software update.
We performed the download of the apps and the filtering from 2023-04-07 to 2023-09-03, so only apps that were online and available in that time frame were further considered.

Once a package is marked for analysis, we download the package from the AndroZoo servers and pass it to our analysis framework \MOBSEC{}, which is described in the next section.

\subsection{Mobsec Analytika}
\MOBSEC{} is a framework for large-scale app analysis.
It is suited both for large-scale (one analysis for many apps) and in-depth (many analyses for one app) experiments.
The design goal of \MOBSEC{} is to be a flexible and extensible framework, as apps can be analyzed in multiple ways and with different algorithms and techniques.
This adaptability is achieved with custom modules, which users can create to extend the functionality of the framework.

For our analysis, we developed four modules for the four different \acp{API} we target.
Modules operate on files extracted from an \ac{APK}, and each module focuses on one specific attribute of the code.
For our experiment, we use \textit{apktool}\footnote{\url{https://apktool.org/}} as a bytecode disassembler to extract information from \ac{APK} files.
The modules then search through the disassembled code and look for \ac{API} function calls for each of the \acp{API} we are interested in.
When such a match is found, information about the match is collected, such as the file path or usage details.
Reports are written to a \acf{JSON} file and can be processed further for evaluation.

If any errors occur during analysis, we exclude the app from the final results.
Errors can be attributed to timeouts in the analysis, when apps are too complex or willingly obfuscated to protect against analysis.
In our analysis (see \autoref{forgotee:analyzing}), we also consider and quantify the apps that failed our analysis.
If no errors occur, every module yields a definitive result, which is either positive (at least one match) or negative (no matches).

After the analysis step, we save the collected data in a database.
The final results, as presented in \autoref{forgotee:results}, are aggregated separately using the \textit{pandas} data analysis library.

\section{Analyzing Apps and \acp{API}} \label{forgotee:analyzing}
As described in the methodology section, our analysis is based on the static analysis of app code.
In this section, we describe which data we collect and how we define \ac{API} matches in Android apps. 

\subsection{Android-specific \ac{API} calls}
In order to detect the presence of \ac{TEE}-related \ac{API} calls, we researched the actual Android classes and interfaces which are used to hold and manage relevant data.
We then also extracted method calls that represent instantiation or usage of these classes, such that unused classes may not trigger a false positive in our app analysis.
The patterns used for matching the \acp{API} are found in \autoref{forgotee:matching-patterns}.

For the KeyStore \ac{API}, we have a total of 23 classes, 15 from the \texttt{android.security.keystore} package and 8 classes from the parent package \texttt{android.security}.
Each class is used with cryptographic key pairs, either in generation, retrieval, or validation.
We also included \texttt{Exception} classes that are used to catch and handle errors, such as \texttt{KeyStoreException}.
From all classes, we included 166 methods that may be used to instantiate objects, retrieve data or perform operations.
During our code analysis, we try to find matches to these methods.

For \ac{DRM} we collected 22 classes and 13 methods.
While most apps will use custom \ac{DRM} classes to handle specific use cases, the app still needs to call methods from the \texttt{android.media.MediaDrm} class and from \texttt{android.drm} packages to use the hardware-based \ac{DRM} protections.

\begin{sloppypar}\tolerance 900
Biometric authentication is also one of the features that work with a \ac{TEE}-based application.
Biometric information such as fingerprints or face scans are considered sensitive information and may not be extracted from a device.
In the best case, this is guaranteed by encrypting the biometric signatures and only decrypting them in the \ac{TEE}.
With Android's StrongBox concept, which includes a physical coprocessor that handles sensitive data and operations, extraction of the data through a remote attack is made even more difficult.
Android offers 8 classes and 20 methods of the \texttt{android.hardware.biometrics} package to initialize and use biometric features.
\end{sloppypar}

Finally, we include Protected Confirmation detection into our analysis.
The feature uses \ac{UI} elements composed in the \ac{TEE} to show a confirmation dialog to the user.
Developers need to consider up to 5 classes with 12 methods to be able to use Protected Confirmation.

\subsection{Cryptographic Libraries} 
Regarding our second research question, we had the hypothesis that some \ac{API} calls may happen inside known security-focused software libraries.
Especially cryptographic libraries are popular and often recommended by security professionals.
In order to verify this hypothesis, we detected usage of crypto libraries in the app.
Similar to the above modules, we included 11 popular libraries, with a total of 37 packages associated with these libraries.
These software libraries expose several \acp{API} that can be called inside the app's source code.
Therefore, we look for imports of classes and methods that reference the unique package name of software libraries to detect usage.
For example, all classes of the popular BouncyCastle library originate in the package \texttt{org.bouncycastle.crypto}, and therefore finding usage of related methods means looking for the name in the app code.
We manually researched and built a list of patterns which were mostly involved in the usage of cryptography for every library, to minimize false positive matches.
This list is found in \autoref{forgotee:matching-patterns}.

Our analysis also includes 9 additional crypto libraries which are included as native libraries, often in the form of static shared object (\texttt{*.so}) files.
These are more difficult to detect, as the relevant methods are often included in the library as compiled code, and not easily analyzed with static analysis tools.
Therefore, we have a second list of potential file names that are commonly used for the specific native libraries.
With a regular expression, we also catch slight variations of these, so a search for \texttt{libraryA} might include the files \texttt{libraryA\_10.2.3.so} and \texttt{libraryA.so.1.2}.
While not perfect, this method provided fast detection with the most amount of correct matches at a low false-positive rate, at the cost of a higher false-negative rate (libraries undetected because of non-matching filename).

\subsection{Match Location} \label{sec:forgotee-match-location}
In our second research questions, we ask about the context of the \ac{API} usage.
It is relevant to know if apps willingly use the target classes and methods, or if they are only used indirectly by third-party libraries shipped with the app.
When a match is found with one of the \ac{API} calls, \MOBSEC{} also saves the file path of the match.
From this path, we can extract the actual package the match was found in, as the paths follow the package naming pattern of Android.
Therefore, the compiled bytecode files from package \texttt{com.example.xyz} are found in the path \texttt{./com/example/xyz/}.

There might be exceptions to this rule, especially when the app code was obfuscated or \textit{minified}.
In most cases, \textit{ProGuard} is used, which until recently was used by the Android build environment to \enquote{shrink} and obfuscate bytecode~\cite{googleShrinkCode}.
ProGuard shortens package names to alphabetical structures (e.g., \texttt{b.d}, \texttt{c.d}), and class names are transformed to single uppercase letter combinations.
For our experiment, this has little significance, as we know the obfuscation rules and can deduce that such shortened packages usually belong to the main application.
We still defined a group of libraries as \enquote{obfuscated} if the package names were found to be invalid according to Android's application ID rules~\cite{androidAppId}.
Moreover, we observed that in the majority of cases, included libraries were \textbf{not} obfuscated or minified.
The reason could be that code minification and shrinking need to be explicitly enabled and configured during compilation, especially for third-party code~\cite{googleShrinkCode}.

\subsection{Further Metadata}
During analysis, several data points were collected that are not directly part of the results, but helped keep the data clean and easy to analyze.
Since our \acp{APK} originate from the third-party service AndroZoo~\cite{DBLP:conf/msr/AllixBKT16}, we perform rudimentary checks to assure their metadata about the apps is correct.
For this, we extract the reported package name of the app and compare it to the one AndroZoo provides.
We also verify the files' SHA-256 hashes to exclude download errors and bit flips during transmission and processing.
Since we also check the existence of the app on the Play Store, we save some data related to it, such as app title, downloads, and author information.

We measure the execution time for every analysis, both for the whole process and individual modules.
In order to get the highest number of apps analyzed in a limited time, we applied a timeout of 15 minutes.
After this time, the analysis is aborted and an error is logged.
If an analysis is successful, the various analysis results include file names and line numbers of the individual matches.
This is done to be able to reproduce the analysis manually or for tracing errors and bugs.
Using this information, we can also deduce if a call is done from the main application or from one of the included third-party software libraries, as required by one of our research questions.

\section{Results} \label{forgotee:results}
The analysis was performed over \experimentDurationDays{} days and a total of \collectedApps{} apps were collected.
We examined \analyzedApps{} \acp{APK}, of which \analyzedFailedApps{} apps failed the analysis.
In this section, we present the results of our data analysis.

\subsection{General Usage}
We successfully analyzed \analyzedSuccessApps{} apps. 
From these, \matchNoTee{} apps (\matchNoTeePercent{}) contain no \ac{API} matches, \matchAtLeastOneTee{} apps (\matchAtLeastOneTeePercent{}) contain references to at least one \ac{TEE}-related \ac{API}, and only \matchAllTee{} contain references to all four \acp{API}.
By excluding Protected Confirmation, which is the least used \ac{API}, we get a total of \matchAllButConfPrompt{} apps (\matchAllButConfPromptPercent{}) that use all the other \acp{API}.
Of all apps that include a match, \matchObfuscatedAppsWithMatchPercent{} of them include at least one obfuscated library (per our definition in \autoref{sec:forgotee-match-location}).

Between the four \acp{API} we looked for, this is the distribution of apps that contain at least one reference to them:
\begin{itemize}
  \item KeyStore: \matchKeystoreTotal{} (\matchKeystoreTotalPercent{})
  \item DRM: \matchDrmTotal{} (\matchDrmTotalPercent{})
  \item Biometrics: \matchBiometricsTotal{} (\matchBiometricsTotalPercent{})
  \item Protected Confirmation: \matchConfPromptTotal{} (\matchConfPromptTotalPercent{})
\end{itemize}

We verified that the package name of the \ac{APK} file matches the name reported by AndroZoo.
In \pkgNotMatching{} cases, the actual package name did not or only partially match the reported name, with \pkgNotMatchingStartswith{} of these at least sharing the same package prefix (e.g., \texttt{com.package} and \texttt{com.package.xyz}).
The remaining apps have probably been mismatched during the data collection by the AndroZoo project.
Since we used the reported package name defined in the \ac{APK}'s manifest file, these errors do not impact our results.

The duration of an analysis depends on the complexity of the app and the size of its code base.
While durations range between \elapsedTimeMin{} and \elapsedTimeMax{} seconds, the median analysis time is \elapsedTimeMedian{} seconds.
If an analysis takes longer than our selected timeout of 900 seconds (15 minutes), we consider the analysis \enquote{failed} and therefore it does not appear in these statistics anymore.

\begin{figure}[htb]
    \centering
    \includegraphics[width=\linewidth]{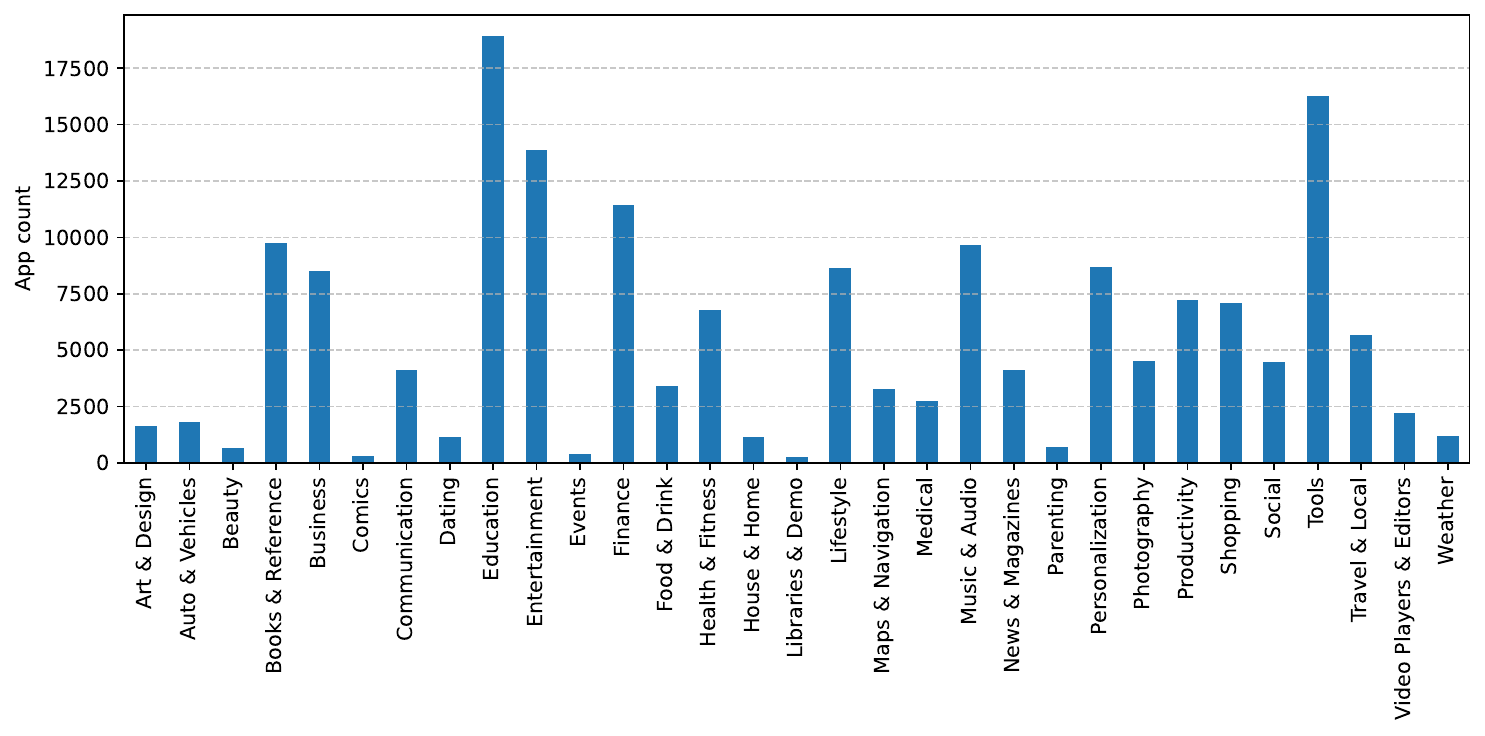}
    \caption{Number of analyzed apps per Play Store category}%
    \label{fig:forgotee-apps-per-category}
\end{figure}

For the quality of our collected sample, we extracted every app's category to make sure that our dataset is varied enough over all types of mobile apps.
By selecting apps at random, we get the category distribution shown in \autoref{fig:forgotee-apps-per-category}.
This extra information in our dataset allows showing differences in \ac{API} usage for different app types, without having to analyze the inner workings and purposes of singular apps.

\begin{figure*}[htb]  %
    \centering
    \includegraphics[width=\linewidth,trim={0 5pt 0 0},clip]{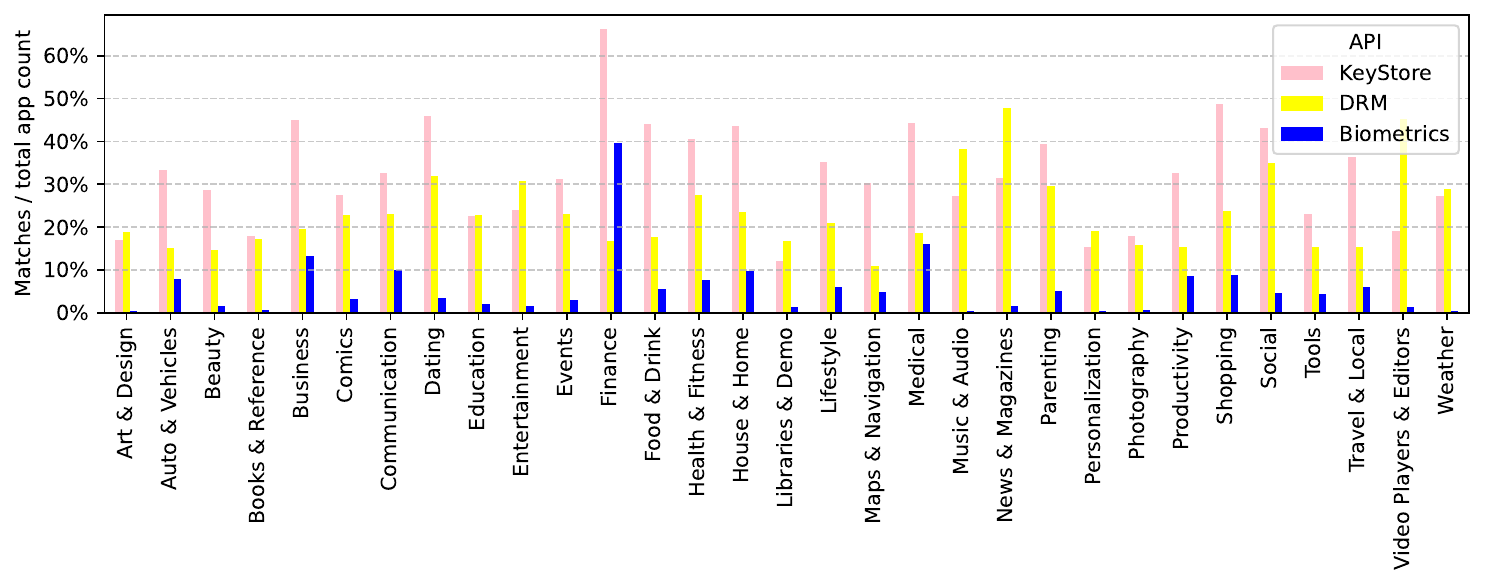}
    \caption{Relative \ac{API} matches per app category. The bars determine how many apps in a category contain calls to a particular \ac{API}, out of all successfully analyzed apps for the category. The Protected Confirmation \ac{API} was left out intentionally for improved layout.}
    \label{fig:forgotee-matches-per-category-relative}
\end{figure*}

Next, we want to see how the actual \acp{API} are distributed over app types.
For this, we combine the category information with the number of apps where a specific \ac{API} was detected.
In \autoref{fig:forgotee-matches-per-category-relative}, we show the relative matches for 3 of the 4 \acp{API} over all analyzed apps.

From the graph, we see that over 65\% of apps in the Finance category contain calls to the KeyStore.
The category also holds the record for biggest share of apps including Biometrics.
For \ac{DRM}, there are peaks in Music \& Audio, News \& Magazine, and Video Players \& Editors.
These results might not be surprising to the reader, but they strengthen our claim of the correctness of our analysis.

For our second research question, we also analyzed the location of these matches and show how the matches are distributed among the main app and any included libraries.
In the following, we present the results for both types of matches.

\subsection{\acp{API} in Libraries}
Our matching method collects the location of the \ac{API} match, which can be used to detect which code file calls the corresponding class or method.
We take a look at \ac{API} matches in included third-party libraries, which we will call \textbf{inlib usage} for the rest of this work.

From all analyzed apps, over \matchInlibApps{} apps (\matchInlibAppsPercent{}) contain inlib usages.
Apps contain an average of \inlibMeanLibraryCount{} (median: \inlibMedianLibraryCount{}) libraries where a match is detected.
Over all matches found in apps, both in libraries and in the main app, around \inlibMatchesPercent{} are attributed to inlib usages.
Therefore, the majority of \ac{API} invocations do not happen as part of the main app, but inside the libraries that are imported by the apps.

Furthermore, compared to the number of apps that contain at least one match, about \matchInlibAppsWithMatchPercent{} of apps were found to contain inlib uses. %
As before, we looked at the \acp{API} that were matched.
From the \matchInlibApps{} apps that include at least one inlib usage, the KeyStore was found in \inlibMatchKeystorePercent{}, \ac{DRM} in \inlibMatchDrmPercent{}, and Biometrics in \inlibMatchBiometricsPercent{} of apps.
For Protected Confirmation, we found \inlibMatchProtconf{} apps (\inlibMatchProtconfPercent{}).

While we cannot work out if a specific call is actually executed during runtime, we can identify the top libraries for every \ac{API} used inside these apps, which we present in the following sections.
In the following, we present the libraries that contain relevant \ac{API} calls.

\subsubsection{KeyStore}
As described in \autoref{sec:forgotee-match-location}, we locate a match inside the app structure and extract the corresponding package name of the library.
For KeyStore, we detected \inlibKeystorePackages{} unique libraries with inlib usage.
The top 10 libraries are shown in \autoref{tbl:forgotee-inlib-keystore-top10}.

\begin{table}[htbp]
    \centering
    \tiny
    \begin{tabular}{|l|c|r|}  \hline
        \textbf{Package} & \textbf{Apps} & \textbf{Purpose} \\  \hline
        com.appsflyer & 6443 & App analytics, event tracking ~\cite{comAppsflyer} \\  \hline
        androidx.biometric & 5124 & Android Jetpack support library ~\cite{androidJetpack} \\  \hline
        com.flurry.sdk & 4717 & App Analytics, Reporting ~\cite{comFlurrySdk} \\  \hline
        androidx.security.crypto & 3824 & Android Jetpack support library ~\cite{androidJetpack} \\  \hline
        com.microsoft.appcenter.utils.crypto & 3593 & Analytics, Reporting, VS App Center SDK ~\cite{comMicrosoftAppcenterUtilsCrypto} \\  \hline
        com.google.crypto.tink.integration.android & 3453 & Tink crypto library ~\cite{googleTink} \\  \hline
        com.amazonaws.internal.keyvaluestore & 1845 & AWS SDK for Android ~\cite{comAmazonawsInternalKeyValuestore} \\  \hline
        com.silkimen.cordovahttp & 1154 & Cordova Advanced HTTP plugin ~\cite{comSilkimenCordovahttp} \\  \hline
        com.oblador.keychain.cipherStorage & 953 & KeyStore Access for React Native apps ~\cite{comObladorKeychainCipherStorage} \\  \hline
        expo.modules.securestore & 916 & Expo SecureStore library ~\cite{expoModulesSecurestore} \\  \hline
    \end{tabular}
    \caption{Top 10 packages by app count that use KeyStore}
    \label{tbl:forgotee-inlib-keystore-top10}
\end{table}

While the use cases for some libraries might be unknown, we also see several support packages that offer easy access to KeyStore functionality, such as Android Jetpack.
From the list of packages, it is apparent that not all app developers may be aware of the use of these \acp{API} in their app.
By manual inspection of some apps that include the Android Jetpack~\cite{androidJetpack} library, we found that the majority of apps are shipped with all or most of the packages offered by the library.
In addition, most apps in our subset did not use or need to use the KeyStore functionality.
We did not find any obfuscation or minimization of the library, which may explain why so many apps \enquote{include} KeyStore without using it.
This observation is also valid for the other \acp{API} in our experiment.

\subsubsection{\acl{DRM}}
For \ac{DRM}, we detected \inlibDrmPackages{} unique libraries that use it.
As \ac{DRM} is a very specific feature of Android, most uses are based on showing protected content to the user.
Again, the top packages that include \ac{DRM} \acp{API} are presented in \autoref{tbl:forgotee-inlib-drm-top10}.

\begin{table}[htbp]
    \centering
    \tiny
    \begin{tabular}{|l|c|r|} \hline
        \textbf{Package} & \textbf{Apps} & \textbf{Purpose} \\ \hline
        com.google.android.exoplayer2.drm & 18558 & ExoPlayer media player~\cite{exoplayer} \\ \hline
        com.applovin.exoplayer2.d & 3471 & Advertising and Monetization~\cite{applovin} \\ \hline
        mono.android.drm & 2410 & DRM for Xamarin .NET Android ~\cite{xamarin} \\ \hline
        androidx.media2.exoplayer.external.drm & 1031 & Android Jetpack media player ~\cite{androidJetpack} \\ \hline
        androidx.media2.player & 1010 & Android Jetpack media player~\cite{androidJetpack} \\ \hline
        android.media & 969 & Android Media API~\cite{androidmediapackage} \\ \hline
        com.mbridge.msdk.playercommon.exoplayer2.drm & 879 & Advertising and Monetization~\cite{mintegral} \\ \hline
        com.segment.analytics & 700 & Segment Analytics~\cite{segment} \\ \hline
        com.google.ads.interactivemedia.v3.internal & 640 & Google Media Ads SDK ~\cite{googlemediaads} \\ \hline
        com.google.android.exoplayer.drm & 574 & ExoPlayer media player~\cite{exoplayer} \\ \hline
    \end{tabular}
    \caption{Top 10 packages by app count that use \acs{DRM}}
    \label{tbl:forgotee-inlib-drm-top10}
\end{table}

In the list of packages, we see plenty of references to \textit{ExoPlayer}, a media player \ac{API} for Android which can be extended~\cite{exoplayer}.
It is included with the Android Jetpack framework, so developers may also use that implementation in their apps.
Since ExoPlayer uses Android's \texttt{MediaDrm} \ac{API} for protected playbacks, most implementations that extend ExoPlayer include code invoking the lower-level Android \acp{API}.
As some ads may also contain copyrighted material, the feature is also prominently used for in-app advertisement libraries, such as {Google Ads} or {AppLovin}.
As a more specialized use case, \ac{DRM} is contained in far fewer packages than KeyStore.

\subsubsection{Biometrics}
The Biometrics \ac{API} was detected in \inlibBiometricsPackages{} unique libraries.
The top packages are shown in \autoref{tbl:forgotee-inlib-biometrics-top10}.

\begin{table}[htbp]
    \centering
    \tiny
    \begin{tabular}{|l|c|r|} \hline
        \textbf{Package} & \textbf{Apps} & \textbf{Purpose} \\ \hline
        androidx.biometric & 8056 & Android Jetpack support library~\cite{androidJetpack} \\ \hline
        android.hardware.biometrics & 559 & Android Biometrics SDK~\cite{androidbiometrics} \\ \hline
        androidx.appcompat.widget & 137 & Android Jetpack support library~\cite{androidJetpack} \\ \hline
        com.exxbrain.android.biometric & 117 & Biometrics library for Android~\cite{exxbrainbiometric} \\ \hline
        org.robolectric.shadows & 80 & Robolectric Unit testing library~\cite{robolectric} \\ \hline
        com.tuya.smart.biometric.finger.compat & 58 & Tuya IoT SDK for Tuya Biometric Readers~\cite{tuyabiometric} \\ \hline
        com.tuya.security.base.finger & 52 & Tuya IoT SDK for Tuya Biometric Readers~\cite{tuyabiometric} \\ \hline
        com.an.biometric & 39 & Biometric Authentication Library~\cite{biometricauthsample} \\ \hline
        com.ts.common.internal.core & 34 & Identity verification~\cite{transmitsecurity} \\ \hline
        com.inmobile & 30 & SMS Marketing~\cite{inmobile} \\ \hline 
    \end{tabular}
    \caption{Top 10 packages by app count that use Biometrics}
    \label{tbl:forgotee-inlib-biometrics-top10}
\end{table}

As an even more specialized security \ac{API}, even less libraries include the functionality for other apps to use.
Therefore, we see most usages coming from the Android Jetpack library or from the \ac{API} package itself.
Most other uncommon uses are situational and involve security checks or the communication with custom fingerprint readers.
For this category, we identified a higher rate of code obfuscation than with the other \acp{API}.
We estimate that over 350 apps include a biometric library, which we could not identify due to obfuscation.
While still below the majority of apps that use known libraries, it is a considerable number of apps that we were unable to identify in this category.

\subsubsection{Protected Confirmation}
As we described in previous sections, we only had \matchConfPromptTotal{} apps containing Protected Confirmation code, and all of these matched in \inlibProtconfPackages{} third-party libraries.
In three apps, the code resided in the \texttt{android.security} package, which is the standard Android \ac{SDK}.
For the remaining app, the \ac{API} call was located inside \texttt{com.android.tools.r8.internal}, which seems related to the Android's default \textit{D8 compiler}~\cite{googleSourceR8}.
As the apps in question do not have any apparent connections to Google or Android, we could not reproduce why this code was included in the app package.

\subsection{Direct \ac{API} Usage}
In this section, we present how apps use the \acp{API} directly inside the main app package, rather than having the code inside a third-party library.
We call this \textbf{inmain usage}.
The difference is significant: In the previous sections, the code to invoke \ac{API} calls is included inside an app, but we cannot reliably tell if it is actually used.
But if we know that developers actually included the code themselves, combined with the way apps are packaged and minified (see \autoref{sec:forgotee-match-location}), we can assume that the \ac{API} is actually being used.

From all analyzed apps that contain at least one match, about \matchInmainAppsWithMatchPercent{} of apps were found to contain inmain uses, and \matchInmainOnlyAppsWithMatchPercent{} having exclusively inmain usage.
As before, we looked at the \acp{API} that were matched.
From the \matchInmainApps{} apps that include at least one inmain usage, the KeyStore was found in \inmainMatchKeystorePercent{}, \ac{DRM} in \inmainMatchDrmPercent{}, and Biometrics in \inmainMatchBiometricsPercent{} of apps.
No matches were found for Protected Confirmation.
By direct comparison, the share of KeyStore matches is significantly higher for inapp usage, while in turn \ac{DRM} and Biometrics is reduced.

\begin{figure}[htbp]
    \centering
    \includegraphics[width=\linewidth,trim={0 5pt 0 0},clip]{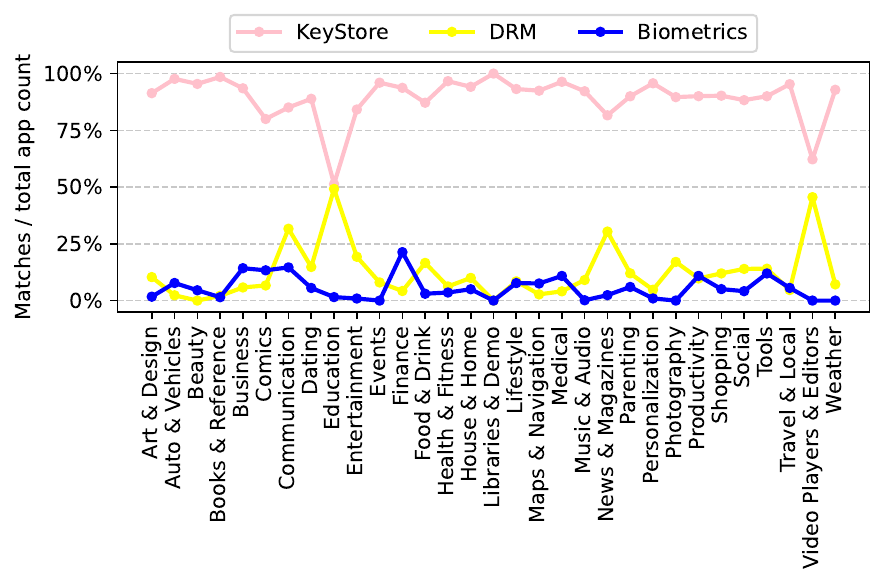}
    \caption{Relative \ac{API} matches per app category for inmain usage.%
    }
    \label{fig:forgotee-inmain-matches-per-category-relative}
\end{figure}

In order to understand how and why developers choose to use the \acp{API} directly in their app, we aggregate the data with the category of the apps.
The result is shown in \autoref{fig:forgotee-inmain-matches-per-category-relative}.
Looking at the differences between categories, we see that Biometrics are sparsely used directly, with a share of below 25\%.
As a specialized use case, \ac{DRM} is again used mainly in the categories \textit{Education} and \textit{Video Players \& Editors}.
Through our analysis, we know that the \ac{API} is also used for showing ads, which may explain the spikes in the categories \textit{Communication} and \textit{News \& Magazines}.
We can also observe another peculiarity of the data in the categories, with a high usage of \ac{DRM} in combination with the KeyStore \ac{API}.
KeyStore is used universally across all categories by developers, its usage mostly ranging between 75\% and 99\%.
When \ac{DRM} has its maximum values, the Keystore seems to have a local minimum.
Since we compare the matches with all apps of a category, and any app can contain multiple \ac{API} matches (but has at least one match), the sum of all values in a category can go over 100\%.
The values are also detached from the actual matching method, as the graph only counts the number of apps that include one or more matches, not the actual number of matches in one app.
Therefore, we can confidently exclude any calculation mistake in our analysis.
Still, there is currently no logical explanation for this phenomenon.

\subsection{Cryptographic Library Usage}

Known by the slogan \enquote{Don't roll your own crypto}, the practice of using ready-made and established third-party libraries for encryption is widely understood among experts to be the safest way to use cryptographic functions.
We detected \matchCryptoImports{} apps (\matchCryptoImportsPercent{} of all analyzed apps) that include at least one of the surveyed crypto libraries.
The libraries with their relative app matches are shown in \autoref{tbl:forgotee-crypto-libs-filled}.

\begin{table}[htbp]
    \centering
    \scriptsize
    \begin{tabular}{|r|r||r|r|}  \hline
    \textbf{Software library}                        & \textbf{\# apps}           & \textbf{Native library}    & \textbf{\# apps}        \\  \hline
    Java Security \cite{javaSESecurity}              & \matchCryptoLibJSE{}       & Crypto++ \cite{cryptopp}   & \matchCryptoCryptoPP{}  \\  \hline
    Android Jetpack \cite{androidJetpack}            & \matchCryptoLibAndX{}      & Botan \cite{botanCrypto}   & \matchCryptoBotan{}     \\  \hline
    BouncyCastle \cite{bouncyCastle}                 & \matchCryptoLibBoCa{}      & Libgcrypt \cite{libgcrypt} & \matchCryptoLibG{}      \\  \hline
    Nimbus JOSE + JWT \cite{joseLibrary}             & \matchCryptoLibNimDS{}     & OpenSSL \cite{openSSL}     & \matchCryptoOpenSSL{}   \\  \hline
    Spring Security JWT \cite{javaSpring}            & \matchCryptoLibSprSec{}    & GnuTLS \cite{gnuTLS}       & \matchCryptoGnuTLS{}    \\  \hline
    Jasypt \cite{jasyptJava}                         & \matchCryptoLibJasypt{}    & Sodium \cite{libsodium}    & \matchCryptoSodium{}    \\  \hline
    JNCryptor \cite{jnCryptor}                       & \matchCryptoLibJNCryptor{} & GNU Nettle \cite{nettle}   & \matchCryptoNettle{}    \\  \hline
    AWS KMS \cite{awsKMS}                            & \matchCryptoLibAWSKMS{}    & wolfSSL \cite{wolfSSL}     & \matchCryptoWolfSSL{}   \\  \hline
    Google Tink \cite{googleTink}                    & \matchCryptoLibGoTink{}    & ~                          &                         \\  \hline
    Apache Commons Crypto \cite{apacheCommonsCrypto} & \matchCryptoLibApaComm{}   & ~                          &                         \\  \hline
    Apache Tuweni \cite{apacheTuweni}                & \matchCryptoLibApaTuw{}    & ~                          &                         \\ \hline \hline
    \textbf{Apps with software library} & \matchCryptoImports{} & \textbf{Apps with native library} & \matchAtLeastOneNativeLib{}      \\ \hline
    \end{tabular}
    \caption{List of cryptographic libraries with the number of apps that include them}
    \label{tbl:forgotee-crypto-libs-filled}
\end{table}

Almost all apps in our dataset include calls to the cryptography classes of the Java Class Library, the set of functions and classes available to all applications based on the Java Virtual Machine.
One of the central interface classes in the library is \texttt{java.security.Key}, which handles several types of cryptographic keys.
It is also the origin for the majority of matches in our experiment.
Apart from the standard library, we can identify \textit{BouncyCastle} and \textit{Google Tink} as the most used general-purpose cryptographic libraries.

Before our analysis, we expected that most of the crypto libraries for Android may offer KeyStore support in their implementations, as it is the only secure way to protect against root-level attackers on mobile devices.
Therefore, we cross-check this assumption by looking at the libraries where KeyStore matches occur, and compare it to our list of cryptographic libraries.
We found that only two libraries, the same we identified in \autoref{tbl:forgotee-inlib-keystore-top10}, include calls to the KeyStore \ac{API}: \texttt{androidx.security} (Android Jetpack) and \texttt{com.google.crypto.tink} (Google Tink).
While both libraries are owned by Google, and Jetpack is becoming part of the official Android development ecosystem, the integration of Android \acp{API} in Tink as a cross-platform cryptographic library is actually a bonus feature for Android users.
For the open-source encryption libraries in our selection, we could verify that they really did not include KeyStore specific code.

For the native libraries, we only found \matchAtLeastOneNativeLib{} apps that contain at least one.
These apps mostly used the \texttt{libssl} and \texttt{libcrypto} libraries by OpenSSL.
As described in \autoref{forgotee:analyzing}, the detection of native libraries is more error-prone, and therefore readers should take these results with caution.

\section{Related Work} \label{forgotee:related-work}

In past research, large-scale static app analysis has been used to various degrees to analyze app behavior, mainly focusing on \enquote{security and privacy issues}~\cite{DBLP:journals/infsof/LiBPRBOKT17}.
Studies have been performed to assess the amount of software reuse in apps~\cite{DBLP:journals/software/RuizANDBH14}, the accessibility of \ac{UI} elements~\cite{DBLP:conf/assets/RossZFW18} or vulnerabilities in mobile games~\cite{DBLP:conf/uss/ZuoL22}.
Third-party cryptographic libraries on mobile devices have been explored and analyzed in the past, and more recent research suggests that 90\% of misuses of the \acp{API} do not come from app developers, but originate in third-party libraries~\cite{DBLP:conf/ccs/0001BD16, DBLP:conf/ccs/MuslukhovBB18}.
In \ac{TEE} security, there has been research that analyzes \ac{TEE} trusted apps~\cite{DBLP:conf/uss/HarrisonVPSG20}, but without considering the software interfaces on the \ac{OS} side.
While we could not find previous works that look at security-focused \acp{API} on Android, there are also similar works to our experimental approach, that we want to mention in the following.

The most adjacent work is SARA, a software library that uses existing native \acp{API} to implement \enquote{remote authorization}~\cite{DBLP:conf/uss/ImranFICB22}.
While the main goal is to present SARA as an alternative to native \acp{API}, the paper also features a large-scale app analysis to determine how apps use \ac{TEE} features.
The work analyzes 112,886 apps from the Play Store, using the same tools as in our experiment (see \autoref{forgotee:methodology}).
Unfortunately, the paper only focuses on key attestation methods and Protected Confirmation, and it reports that only 5 apps were found to include key attestation (a feature of KeyStore) and no apps were found to use Protected Confirmation.
In a usability study performed by the authors, they also found that developers were unable to use the KeyStore and Protected Confirmation \acp{API},
declaring the official documentation was \enquote{complex}.

In ~\cite{DBLP:conf/icissp/HeidTHS22}, the authors collect the top 1,000 Play Store apps to detect and analyze file system accesses in Android's Filesystem \ac{API}.
As a difference to our approach, the work applies a dynamic analysis using \ac{UI} automation to trigger app behavior and the Frida framework to hook the Java \ac{API}.
While they also analyze KeyStore app interactions, the paper does not present any results specific to it.

\begin{sloppypar}\tolerance 900
The paper BrokenFinger~\cite{DBLP:conf/ndss/BianchiFMKVCL18} focuses on the \texttt{keymaster} and \texttt{fingerprintd} \ac{TEE} services.
They survey the usage of the Fingerprint \ac{API} on Android, a predecessor of the Biometrics \ac{API}, by statically analyzing bytecode of 501 apps, selected from a dataset of 30,549 top apps from the Play Store.
They found that 14.37\% of analyzed apps did not contain related \acp{API} calls in their source code, and over 50\% of apps do not use Android's safeguards that verify if the fingerprint sensor was actually touched.
Given the difference in the selection strategy for the dataset, the actual distribution of apps using Biometrics might not be comparable to our experiment.
Their dataset contains 1.4\% of apps that use fingerprint features, while our experiment with over 5 times more apps selected by install count yielded a share of \matchBiometricsTotalPercent{} for Biometrics usage.
\end{sloppypar}

In SafetyNOT~\cite{DBLP:conf/mobisys/IbrahimIB21}, the authors analyze the use of Google SafetyNet, a predecessor of the Play Integrity \ac{API} which offers developers to remotely check the devices their apps run on.
The \ac{API} offers a remote attestation service, where developers can assess the integrity of a user device, with attributes such as bootloader status, emulator detection or modified \ac{OS} package.
The authors collected 163,773 apps, did an automated string search for \enquote{safetynet} on the smali code, which found 19,834 apps containing the keyword, and further analyzed these apps dynamically using the Frida framework.
In the end, they identified 62 apps that contained SafetyNet \ac{API} invocations, and none of them correctly implemented the security checks provided by the service.

\section{Conclusion} \label{forgotee:conclusion}
This work presents a comprehensive analysis of the usage of \ac{TEE}-based \acp{API} for \analyzedApps{} Android apps.
The results of our experiment offer valuable insights into the current landscape of mobile security practices.
For this, we developed \MOBSEC{} as a tool for large-scale app analysis, which we plan to release to the public as an open-source research tool.

Our findings reveal that \ac{TEE} features are mostly used indirectly through the use of third-party libraries.
While developers may be able to directly call the \acp{API}, only in \matchInmainAppsPercent{} of all apps this occurred.
Most apps that use the examined features may use it through the Android Jetpack library, which wraps many functions of the Android \ac{SDK}.
We also learned that Protected Confirmation, as a recent \ac{TEE}-based \ac{UI} protection, is virtually unused by apps.
For KeyStore, we expected to see more usage in popular cryptographic libraries, but only Android Jetpack and Google Tink were found to use the secure KeyStore \ac{API}.

Despite the extensive nature of our study, there exist certain limitations that should be considered when interpreting the results:

\begin{itemize}
    \item Our analysis is restricted to free apps only, excluding paid applications from the study.
    While this limits the impact of our results to a partial view of the overall app ecosystem, almost 97\% of apps on the Play Store are free to download~\cite{42mattersGooglePlay}. %

    \item In our analysis, we did not account for dependencies between \acp{API}.
    A method that is consistently called within another \ac{API} implementation results in matches for two distinct \acp{API}.
    An example of this is Protected Confirmation, which needs a KeyStore-protected key to sign the confirmation data.

    \item We did not investigate system apps in this study.
    System apps may contain a higher degree of \ac{TEE} usage within the system, as they are usually more privileged than third-party apps.
    Unfortunately, we were unable to obtain enough samples from different devices and manufacturers to include these apps in our analysis.
    Therefore, we limited the scope to third-party apps available to most devices and users.

    \item Code obfuscation poses a challenge for accurate static analysis.
    While most manually analyzed apps were not heavily minified or compressed, the presence of obfuscated code in some applications may have affected the reliability and precision of our analysis, potentially leading to false-negatives, such as missing matches or undetected libraries.
\end{itemize}

Our research fills a significant gap in the field of mobile app research, but there are more promising opportunities for future research.
One direction worth exploring is the incorporation of a user study involving mobile app developers.
By directly engaging with practitioners, we might gain a more in-depth understanding of what developers know about app security and \acp{TEE}, and how their experience with implementing or including security features in their apps are.
Some minor studies have been performed for related work, but a larger study might yield more insights into the actual usage of Android \acp{API}.
We trust that our work will be instrumental in advancing both \ac{TEE} research and the overall landscape of mobile security in a more user-centric direction.

\section*{Availability}
We publish the source code of \MOBSEC{}, as well as a list of all analyzed apps and any research artifacts (e.g., \MOBSEC{} outputs).
Because of legal concerns, we do not publish the collected \ac{APK} files, but provide SHA-256 hashes of all analyzed files.
The repository can be found here: \url{https://www.cs1.tf.fau.de/android-tee-study/}  

\bibliographystyle{plain}
\bibliography{references}

\appendix

\clearpage

\counterwithin{table}{section}

\section{Matching patterns} \label{forgotee:matching-patterns}

\begin{table}[H]
    \centering
    \tblsizeappendix
    \csvreader[tabular=|l|,
        respect dollar,
        table head=\hline \bfseries Pattern \\\hline,
        late after line=\\\hline]%
        {include/data/keystore-patterns-shortened.csv.txt}{pattern=\pattern}%
        {\pattern }%
        \caption{Partial list of patterns for KeyStore detection}
    \label{tbl:forgotee-keystore-patterns}
\end{table}

\begin{table}[H]
    \centering
    \tblsizeappendix
    \csvreader[tabular=|l|,
        respect dollar,
        table head=\hline \bfseries Pattern \\\hline,
        late after line=\\\hline]%
        {include/data/drm-patterns.csv.txt}{pattern=\pattern}%
        {\pattern }%
        \caption{Patterns for DRM detection}
    \label{tbl:forgotee-drm-patterns}
\end{table}
 
\begin{table}[H]
    \centering
    \tblsizeappendix
    \csvreader[tabular=|l|,
        respect dollar,
        table head=\hline \bfseries Pattern \\\hline,
        late after line=\\\hline]%
        {include/data/biometric-patterns.csv.txt}{pattern=\pattern}%
        {\pattern }%
        \caption{Patterns for Biometric usage detection}
    \label{tbl:forgotee-biometric-patterns}
\end{table}
 
\begin{table}[H]
    \centering
    \tblsizeappendix
    \csvreader[tabular=|l|l|,
        respect dollar,
        table head=\hline\bfseries Library & \bfseries Pattern \\\hline,
        late after line=\\\hline]%
        {include/data/crypto-patterns.csv.txt}{library=\library,pattern=\pattern}%
        {\library & \pattern }%
        \caption{Patterns for cryptography library detection}
    \label{tbl:forgotee-crypto-patterns}
\end{table}

\begin{table}[H]
    \centering
    \tblsizeappendix
    \csvreader[tabular=|l|,
        respect dollar,
        table head=\hline \bfseries Pattern \\\hline,
        late after line=\\\hline]%
        {include/data/confprompt-patterns.csv.txt}{pattern=\pattern}%
        {\pattern }%
        \caption{Patterns for Protected Confirmation detection}
    \label{tbl:forgotee-confprompt-patterns}
\end{table}

\end{document}